\documentclass[aip,amsmath,amssymb,reprint,Conference Proceedings]{revtex4-2}
\usepackage{graphicx}
\usepackage[dvipsnames]{xcolor}
\usepackage{dcolumn}
\usepackage{bm}
\usepackage[utf8]{inputenc}
\usepackage[T1]{fontenc}
\usepackage{mathptmx}
\usepackage{etoolbox}
\usepackage{lmodern}
\usepackage{ulem}
\usepackage[colorlinks,linkcolor=black,citecolor=black,filecolor=black,urlcolor=Blue,bookmarksopen=false]{hyperref}
\usepackage{lipsum}

\makeatletter
\def\@email#1#2{%
 \endgroup
 \patchcmd{\titleblock@produce}
  {\frontmatter@RRAPformat}
  {\frontmatter@RRAPformat{\produce@RRAP{*#1\href{mailto:#2}{#2}}}\frontmatter@RRAPformat}
  {}{}
}%
\makeatother

\begin{document}

\title[High-Field Heat Capacity of Atacamite Cu$_2$Cl(OH)$_3$ for $\mathbf{H} \parallel b$ axis]{High-Field Heat Capacity of Atacamite Cu$_2$Cl(OH)$_3$ for $\mathbf{H} \parallel b$ axis: \\ From Antiferromagnetic Order to Correlated Disorder}

\author{L. St\"{o}dter}
\email{l.stoedter@fz-juelich.de}
\affiliation{J\"{u}lich Center for Neutron Science (JCNS) at Heinz Maier-Leibnitz Zentrum (MLZ), Forschungszentrum J\"ulich GmbH, 85748 Garching, Germany}
\author{A. Demuer}
\affiliation{Universit\'{e} Grenoble Alpes, INSA Toulouse, Université Toulouse Paul Sabatier, CNRS, LNCMI, 38000 Grenoble, France}
\author{T. Kotte}
\affiliation{Hochfeld-Magnetlabor Dresden (HLD-EMFL), Helmholtz-Zentrum Dresden-Rossendorf, 01328 Dresden, Germany}
\author{A. U. B. Wolter}
\affiliation{Leibniz Institute for Solid State and Materials Research IFW Dresden, 01069 Dresden, Germany}
\author{B. B\"uchner}
\affiliation{Leibniz Institute for Solid State and Materials Research IFW Dresden, 01069 Dresden, Germany}
\affiliation{Institut f\"ur Festk\"orper- und Materialphysik and W\"urzburg-Dresden Cluster of Excellence ctd.qmat, Technische Universit\"at Dresden, 01062 Dresden, Germany}
\author{K. C. Rule}
\affiliation{Australian Nuclear Science and Technology Organisation, Lucas Heights, NSW 2234, Australia}
\author{S. S\"{u}llow}
\affiliation{Institut f\"{u}r Physik der Kondensierten Materie, Technische Universit\"{a}t Braunschweig, 38106 Braunschweig, Germany}

\date{\today}

\begin{abstract}
We present the heat capacity of the frustrated sawtooth-chain compound atacamite Cu$_2$Cl(OH)$_3$ in high magnetic fields up to 35\,T for $\mathbf{H} \parallel b$ axis. We clearly observe the change from antiferromagnetic long-range order, indicated by a sharp transition anomaly shifted to lower temperatures when increasing field, to a field-induced spin-liquid-like state with the absence of any transition anomaly. Here, in highest magnetic fields above 32\,T a broad anomaly is present which is shifted to higher temperatures as the field is increased. The low-temperature heat capacity in the high-field region can be described well by an exponential behavior indicating a thermal activation over a spin gap. The findings are consistent with our scenario for the high-field state in atacamite for $\mathbf{H} \parallel c$ axis, where we proposed the sawtooth chains to decouple into two spin subsystems, basal and apical spin sites, leading to a partially polarized state with $M \sim M_\mathrm{sat}/2$.
\end{abstract}

\maketitle

\section{Introduction}

The antiferromagnetic (AFM) quantum sawtooth chain is a fundamental model in low-dimensional frustrated magnetism. Over the years, this spin lattice model has been found to host various unconventional phenomena such as dimerized ground states as well as localized magnon states in applied magnetic fields leading to magnetization jumps and plateaus~\cite{Kubo1993,Schulenburg2002,Blundell2003,Richter2004,Zhitomirsky2004,Derzhko2015,Metavitsiadis2020,Rausch2024}. The sawtooth chain is described by the Hamiltonian

\begin{align}
\mathcal{H}=&\sum_{i~\mathrm{odd}}J\mathbf{S}_{i}\cdot\mathbf{S}_{i+2}+J'\big(\mathbf{S}_{i}\cdot\mathbf{S}_{i+1}+\mathbf{S}_{i+1}\cdot\mathbf{S}_{i+2}\big) \nonumber \\
& -h\sum_i {S}^z_{i}
\label{eq:Hamiltonian_sawtoothchain}
\end{align}

\noindent where $\mathbf{S}_{i}$ is the spin-$1/2$ operator at site $i$, $J$ describes the exchange interaction between basal spins along the backbone of the chain, and $J'$ is the exchange interaction between basal ($i$ odd) and apical spins ($i$ even). The term $-h\sum_i {S}^z_{i}$ describes a magnetic field with Zeeman energy $h = g\mu_\mathrm{B}\mu_0H$. When moving away from the case of the uniform sawtooth chain ($J^\prime=J$) the coupling ratio $J^\prime/J$ changes the ground state from gapped, dimerized for $0.654 \lesssim J^\prime/J \lesssim 2.052$ to quasi-canted for $J^\prime/J < 0.654$ and to quasi-Néel for $J^\prime/J > 2.052$~\cite{Blundell2003,Jiang2015}.

Model compounds of the quantum sawtooth chain are rare: In the past, only the spin-$1/2$ delafossite YCuO$_{2.5}$ and the mineral euchroite Cu$_2$(AsO$_4$)(OH)$\cdot$3\,H$_2$O were discussed in the context of sawtooth-chain physics~\cite{Blundell2003,Kikuchi2011}. For euchroite, which has three inequivalent exchange paths in each sawtooth triangle, a dimerized ground state was found~\cite{Kikuchi2011}, likely with the singlet dimers on one basal-apical bond being stabilized due to the lifted degeneracy of the ground state. For YCuO$_{2.5}$, despite the absence of magnetic order down to the lowest experimental temperature~\cite{Blundell2003}, a ferromagnetic intrachain interaction between third neighbors was found in Ref.~\cite{LeBacq2005} leading to a more complex motif of exchange couplings.

\begin{figure*}[t!]
\includegraphics[width=\linewidth]{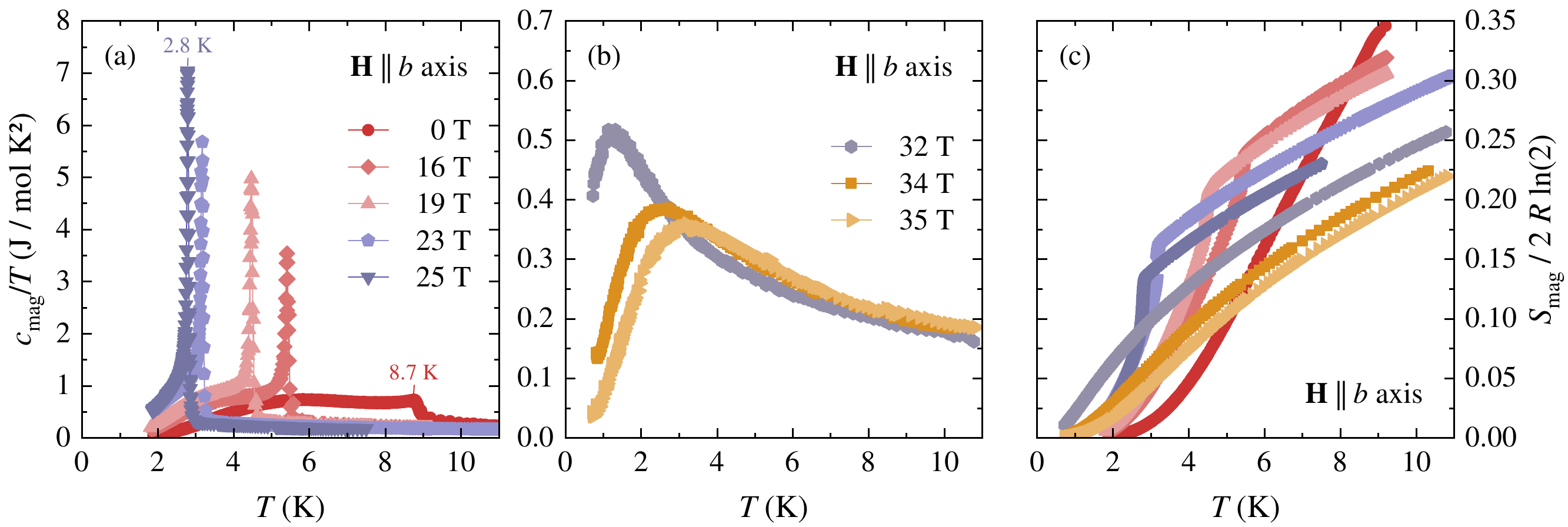}
\caption{(a)--(b) Magnetic specific heat of atacamite, plotted as $c_\mathrm{mag}/T$, measured in fields up to 35\,T ($\mathbf{H} \parallel b$ axis) for the field regions (a) of the AFM phase (together with the zero-field reference) and (b) beyond the AFM phase. (c) Magnetic entropy, calculated from the applied-field heat-capacity data, as function of temperature.}
\label{fig:cmag-Smag-overview}
\end{figure*}

Recently, we have established the copper mineral atacamite Cu$_2$Cl(OH)$_3$, subject of the present study, as spin-$1/2$ sawtooth-chain material~\cite{Heinze2018,Heinze2021,Heinze2025}. With its AFM ground state below $T_\mathrm{N} = 8.9(1)$\,K [magnetic propagation vector $\mathbf{q}_\mathrm{m} = (0.5, 0, 0.5)$~\cite{Heinze2018}], atacamite is clearly different from the materials above. Its sawtooth-chain motif of Cu(1) basal sites and Cu(2) apical sites is embedded into an orthorhombic crystal structure of space group $Pnma$ (No.~62) with $a = 6.030$\AA, $b = 6.865$\,\AA~and $c = 9.120$\,\AA~\cite{Parise1986}. By means of first-principles calculations, the dominant exchange interactions $J_4 \equiv J = 336$\,K and $J_3 \equiv J^\prime = 102$\,K were found to form the AFM sawtooth chains together with far weaker residual couplings~\cite{Heinze2021}. Remarkable about the magnetic properties of atacamite is its behavior in high magnetic fields: Both for $\mathbf{H} \parallel b,c$ axes, we found a plateaulike magnetization at $M \sim M_\mathrm{sat}/2$~\cite{Heinze2021,Heinze2025}, and a combined study of the macroscopic and microscopic properties in high magnetic fields $\mathbf{H} \parallel c$ axis suggested an (effective) dimensional reduction of the spin system~\cite{Heinze2025}. It sets in at the field-induced quantum critical point (QCP) separating long-range AFM order from correlated disorder. In this work, we investigate the high-field heat capacity of atacamite for $\mathbf{H} \parallel b$ axis, this way complementing our earlier studies and gaining new insights into the magnetic phase diagram. From the thermodynamic behavior, we confirm that a quantum phase transition to correlated disorder is also present for this field direction.

\section{Experimental details}

The heat-capacity measurements were carried out at the Laboratoire National des Champs Magn\'{e}tiques Intenses (LNCMI) of the Centre National de la Recherche Scientifique (CNRS), France, using a 36\,T resistive magnet equipped with a $^4$He cryostat. In an additional experiment, a $^3$He refrigerator was used allowing experimental temperatures down to 0.7\,K. We recorded the heat-capacity data using a thermal relaxation technique. Here, a bare Cernox cryogenic temperature sensor served as thermometer, heater and sample holder. We mounted the atacamite sample III-P08, a single crystal cut thin with dimensions of approx.~1.4\,mm $\times$ 1.2\,mm $\times$ 0.3\,mm and a mass of 2.25\,mg (see Ref.~\cite{Heinze2025}), on the back of the Cernox chip using Apiezon grease. Typically, we performed long thermal relaxations with an increase of the the sample temperature being up to 100\% of the bath temperature, which allowed to cover an extended temperature range during one relaxation. The heat capacity was evaluated from the slope of the temperature as function of time~\cite{Wang2001}. We characterized the addenda of the system (heat capacity of the system without sample) by recording the heat capacity of an empty sample holder in zero magnetic field, and we subtracted the addenda heat capacity from the data recorded for the sample.

During the first measurements at highest magnetic fields ($\mathbf{H} \parallel b$ axis), the sample experienced a strong magnetic torque which tore apart the thermometer when the sample was forced into the configuration $\mathbf{H} \parallel c$ axis. For this reason, only fields up to 25\,T were applied for $\mathbf{H} \parallel b$ axis using the newly wired sample holder. For the subsequent experiment with the $^3$He refrigerator, in order to reduce the torque of the sample in highest magnetic fields, we cut a $\sim 200$\,{\textmu}m wide stripe of sample III-P08 (246\,{\textmu}g), referred to as sample III-P08-1. To obtain the magnetic specific heat $c_\mathrm{mag}$, the phononic contribution to the total heat capacity of atacamite, modeled previously in Ref.~\cite{Heinze2025}, was further subtracted from the addenda-corrected data. The resulting specific-heat data are plotted as $c_\mathrm{mag}(T)/T$ in Figs.~\ref{fig:cmag-Smag-overview}(a)--(b).

\section{Results}

In accordance with earlier results~\cite{Heinze2021}, up to 31.5(1)\,T we observe a maximum in specific heat that denotes the transition into the AFM phase [Fig.~\ref{fig:cmag-Smag-overview}(a)]. The transition anomaly sharpens enormously with increasing magnetic field; the maximal specific heat in 25\,T is $\sim 7$\,J mol$^{-1}$ K$^{-2}$ at 2.8\,K. The maximum of $c_\mathrm{mag}/T$ is shifted from 8.7\,K in zero magnetic field down to 2.8\,K in 25\,T, matching well the magnetic phase diagram established in Ref.~\cite{Heinze2021}. The sharp transition in high fields resembles a first-order phase transition.

Notably, the behavior of the specific heat changes for magnetic fields above $32$\,T, i.e., in the region beyond the AFM ground-state phase. Here, $c_\mathrm{mag}/T$ features a broad anomaly with the maximum shifting to higher temperatures as the magnetic field is increased towards the highest experimental field of 35\,T [Fig.~\ref{fig:cmag-Smag-overview}(b)]. No sharp transition anomaly is present down to lowest temperatures indicating the absence of long-range magnetic order. The broad anomaly is reminiscent of a behavior produced by low-lying spin excitations with a field-dependent energy gap. We have recently observed a very similar behavior for $\mathbf{H} \parallel c$ axis in atacamite going together with the suppression of long-range magnetic order, as evidenced by $^1$H-NMR, although far from full field polarization~\cite{Heinze2025}. Also for $\mathbf{H} \parallel b$ axis, the bulk magnetization of atacamite is plateaulike and close to half-saturation above 31.5\,T~\cite{Heinze2021}, which implies an unusual field-induced disordered state with strong spin correlations~[Fig.~\ref{fig:cmag-Smag-overview}(b)].

\begin{figure}[b!]
\includegraphics[width=0.9\linewidth]{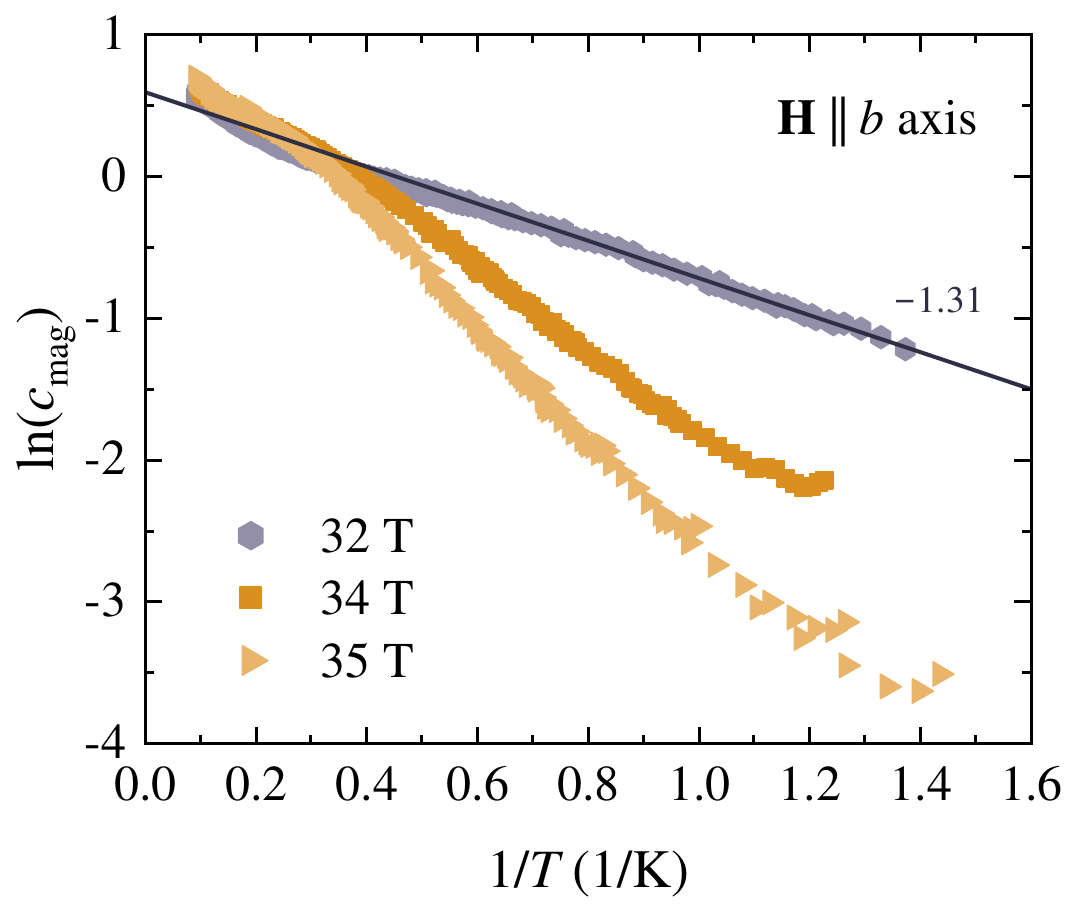}
\caption{Specific heat, plotted as $\ln(c_\mathrm{mag})$ over $1/T$, for the data recorded in the field region beyond the AFM phase. The line denotes a linear fit to the low-temperature behavior of the 32\,T-data resulting in a slope of $-1.31$.}
\label{fig:lncmag-vs-1overT}
\end{figure}

From our data, we have calculated the magnetic entropy $S_\mathrm{mag}$ for the different field points via an integration of the temperature-dependent specific heat

\begin{equation}
S_\mathrm{mag}(T) = \int_0^T \frac{c_\mathrm{mag}(T^\prime)}{T^\prime}\,dT^\prime
\label{eq:magneticentropy}
\end{equation}

\noindent with the $c_\mathrm{mag}$-data extrapolated for $T \to 0$\,K. The resulting magnetic entropy is plotted in Fig.~\ref{fig:cmag-Smag-overview}(c). The zero-field entropy at $T_\mathrm{N}$ is only $\sim 1/3$ of $2R\ln(2)$, the value expected for two spin-$1/2$ per formula unit. Such a behavior is typical for frustrated and/or low-dimensional spin systems where due to persisting short-range spin correlations the entropy is distributed over a larger temperature range defined by the exchange constants~\cite{comm:2Rln2}. 

The entropy data in Fig.~\ref{fig:cmag-Smag-overview}(c) indicate magnetocaloric cooling in the AFM phase, most pronounced for an adiabatic magnetization around 5\,K. They further indicate magnetocaloric heating for an adiabatic magnetization outside the AFM phase. Such a change of the magnetocaloric properties is typical for the presence of a field-induced QCP at the low-temperature end of the AFM phase boundary. We have recently found an analogous behavior in atacamite for $\mathbf{H} \parallel c$ axis with a field-induced QCP at the low-temperature endpoint of the AFM phase at 21.9(1)\,T. The large anisotropy of the critical field, 21.9\,T compared to 31.5\,T, cannot be explained by a possible $g$-factor anisotropy alone. We believe that rather (the combination with) Dzyaloshinskii-Moriya interaction is responsible for this anisotropy of the critical field to suppress the AFM ground-state phase.

To further analyze the low-temperature behavior of the heat capacity, we have plotted in Fig.~\ref{fig:lncmag-vs-1overT} the logarithm of the magnetic specific heat over the inverse temperature. The 32\,T-data, recorded closest to the QCP at 31.5(1)\,T, fall on a line. This reveals that the behavior can be well described using a single exponential function of the form

\begin{equation}
c_\mathrm{mag} \propto \exp(-\frac{\Delta_\mathrm{PM}}{k_\mathrm{B}T}),
\label{eq:cmag-exponentialfit}
\end{equation}

\noindent which approximates a thermal-activation behavior over an energy gap in the spin-excitation spectrum. In Eq.~(\ref{eq:cmag-exponentialfit}), $\Delta_\mathrm{PM}$ is the gap parameter in the quantum paramagnetic regime and $k_\mathrm{B}$ is the Boltzmann constant. We also found such a thermal-activation behavior of the heat capacity previously in atacamite for $\mathbf{H} \parallel c$ axis~\cite{Heinze2025}. For $\mathbf{H} \parallel b$ axis, there are some deviations from a linear behavior at lowest experimental temperatures for the data recorded at 34\,T and 35\,T (Fig.~\ref{eq:cmag-exponentialfit}) where the heat-capacity signal is small and where the individual relaxation processes can be seen from the logarithmic plot. Still, the data suggest an increase of $\Delta_\mathrm{PM}$ as the magnetic field is increased. The gap parameter extracted from the linear fit to the 32\,T-data in Fig.~\ref{fig:lncmag-vs-1overT}, $\Delta_\mathrm{PM}/k_\mathrm{B} = 1.31$\,K, is somewhat smaller than the one we obtained earlier in proximity of the QCP for $\mathbf{H} \parallel c$ axis, $\Delta_\mathrm{PM}/k_\mathrm{B} = 1.78$\,K (23\,T > $\mu_0H_\mathrm{c} = 21.9$\,T)~\cite{Heinze2025}.

\section{Discussion}

The behavior of the high-field heat capacity for $\mathbf{H} \parallel b$ axis closely resembles the behavior we found previously for $\mathbf{H} \parallel c$ axis~\cite{Heinze2025}. In that earlier study, we suggested that for magnetic fields beyond the QCP the sawtooth chains effectively decouple into AFM spin-$1/2$ chains, i.e., the backbones of the sawtooth chains formed by the basal sites, and the remaining field-polarized apical spins producing a bulk magnetization $M \sim M_\mathrm{sat}/2$. The findings for $\mathbf{H} \parallel b$ axis presented in this work are in full agreement with this scenario: The specific heat of atacamite for $\mu_0H > 31.5$\,T would be composed of the specific heat of the two spin subsystems, basal and apical spin sites, but where the signal of the unfrustrated AFM spin-$1/2$ chains would not contribute significantly on the temperature range studied given the large $J$ in atacamite (see Supplemental Material of Ref.~\cite{Heinze2025}). In an early study, we suggested the plateaulike behavior of the magnetization for $\mathbf{H} \parallel b$ axis to be caused by two ordered spin subsystems~\cite{Heinze2021}. The present study rules out this scenario and underlines the quantum nature of the high-field state.

Recent inelastic neutron scattering results provided new insights into the weak interchain network connecting the sawtooth chains. Here, using linear spin wave theory (LSWT), the low-energy spin excitations were modeled as spin waves of AFM zigzag chains ($J_1 = 11.4$\,K, $J_{11b} = 22.5$\,K), formed by the Cu(2) sites along the $a$ axis, in a staggered mean field describing the Cu(1) sites~\cite{Allen2026}. These results prompted that the zigzag chains, although coupled far more weakly than the frustrated sawtooth chains, are responsible for the details of the AFM ordering with $\mathbf{q}_\mathrm{m} = (0.5, 0, 0.5)$ at $T_\mathrm{N}$. Now, using the exchange interactions $J_1$ and $J_{11b}$ obtained from the LSWT analysis, the bare zigzag chains would have a saturation field of $\sim 42$\,T [see Eq.~(2) in Ref.~\cite{Maeshima2003}], in the same range as the critical field in atacamite. This observation supports that the quantum phase transition between AFM order and correlated disorder coincides with the field polarization of the residual zigzag Cu(2) chains. It causes the effective decoupling of the two spin subsystems. The thermal-activation behavior seen in the specific heat (Fig.~\ref{fig:lncmag-vs-1overT}) is consistent with this scenario for the Cu(2) subsystem indicating the opening of a spin gap.

\section{Summary}

In result, atacamite exemplifies a frustrated quantum magnet with a field-induced quantum phase transition from AFM long-range order to a strongly correlated, spin-liquid-like state which occurs due to the weak residual exchange couplings. In this study, we traced the thermodynamic behavior of this material when going beyond the QCP by measuring the heat capacity in high magnetic fields $\mathbf{H} \parallel b$ axis up to 35\,T. The results align with earlier findings for $\mathbf{H} \parallel c$ axis~\cite{Heinze2025}. While atacamite does not represent a model compound of the bare quantum sawtooth chain in the field region of the AFM ground-state phase, the dimensionality of the quantum spin system is effectively lowered upon the suppression of long-range order. The QCP is far from field polarization of the whole system for both $\mathbf{H} \parallel b, c$ axes~\cite{Heinze2021,Heinze2025}, which should only be reached in field energies $h \sim 2J$, corresponding to magnetic fields $\mu_0H \sim 500$\,T for $J=336$\,K. In the future, it would be interesting to trace the bulk magnetization of atacamite to ultrahigh magnetic fields to verify the magnetization behavior of the low-dimensional spin system.

\begin{acknowledgments}
We acknowledge the support of the LNCMI-CNRS and the HLD-HZDR, members of the European Magnetic Field Laboratory (EMFL). This work was partially supported by the DFG under Contract No.~SU229/9-2, SFB 1143 (Project No.~247310070) and Project No.~390858490 (Würzburg-Dresden Cluster of Excellence on Complexity, Topology and Dynamics in Quantum Matter--ctd.qmat, EXC 2147). We thank G. Paskalis and J. McAllister for supplying us with atacamite crystal used for this study. We further thank T. Lampe for the support during the sample preparation.
\end{acknowledgments}

\section*{Data Availability Statement}

The data that support the findings of this study are available from the corresponding author upon reasonable request.


\begin{thebibliography}{99}
\bibitem{comm:2Rln2} Here, it is usually difficult to separate that contribution to the magnetic heat capacity from the lattice contribution at higher temperatures.
\end{thebibliography}
\end{document}